\documentclass[11pt,a4paper]{article}
\pdfoutput=1
\usepackage{jheppub}

\title{A study of the corrections to factorization in $\bar{B}^0 \to D^{* +} \omega \pi^-$}
\author[a,b,c]{S.I.~Eidelman,}
\author[a,b]{L.V.~Kardapoltsev}
\author[a,b,c,1]{and D.V.~Matvienko \note{corresponding author}}
\affiliation[a]{Budker Institute of Nuclear Physics, SB RAS,\\
11, Lavrentieva prospect, Novosibirsk, Russia}
\affiliation[b]{Novosibirsk State University, \\
2, Pirogova street, Novosibirsk, Russia}
\affiliation[c]{Lebedev Physical Institute RAS, \\
53, Leninskiy Prospekt, Moscow, Russia}
\emailAdd{s.i.eidelman@inp.nsk.su}
\emailAdd{l.v.kardapoltsev@inp.nsk.su}
\emailAdd{d.v.matvienko@inp.nsk.su}

\abstract{A factorization hypothesis is tested by examining a form factor 
of the 
$\omega\pi$ production in hadronic $B^0 \to D^{\ast \pm}\omega\pi^\mp$ decays. 
The form factor is compared to that from available $\tau$-lepton 
as well as $e^+e^-$ data using the conserved vector current hypothesis.
The difference of normalizations of form factor shapes from $B$ and 
$\tau~(e^+e^-)$ data indicates the important role of the large $N_c$ limit 
in QCD. Moreover, the growth of the difference between the form factors with 
the $\omega\pi$ invariant mass is related to the perturbative QCD 
corrections of factorization. The current precision of $B$ data does not
allow one to find any evidence of corrections to factorization. A 
promising study could be performed with the Belle II and LHCb data sets.}


\arxivnumber{}

\begin{document}
\maketitle
\flushbottom

\section{Introduction}

Non-leptonic decays of $B$ mesons are usually considered through factorization 
approximation where decay amplitudes are factorized into products of hadronic 
matrix elements of color-singlet currents. Such approximation is not exact 
because only interaction between quarks in the same hadron is taken into 
account without consideration of gluon effects 
which redistribute the quarks. Corrections to factorization are, thus, 
important and could be extracted experimentally. 

Hadronic decays of the $B$ meson are dominated by $B \to D^{(*)} X$ transitions,
where $X$ denotes a system of one, two or more pions. An experimental 
factorization test suggested in ref.~\cite{ligeti} studies the 
correction to factorization 
as a function of the invariant mass of the system $X$ with two or more pions. 
In this paper we focus on the subset of the system $X$ consisting of the
$\omega$ and $\pi$ mesons.

\begin{figure*}[htb!]
\begin{center}
\begin{tabular}{c c}
\includegraphics[width=0.45\textwidth]{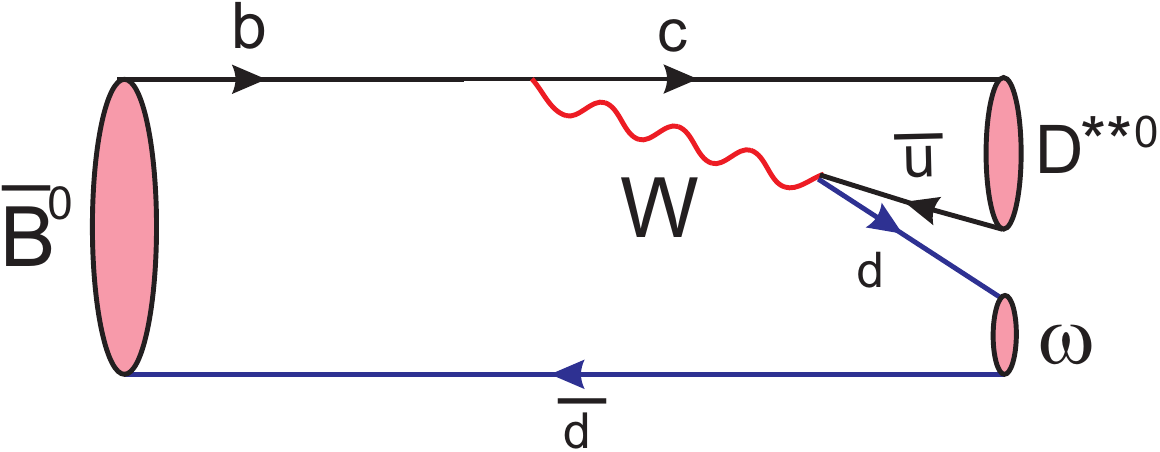} &
\includegraphics[width=0.45\textwidth]{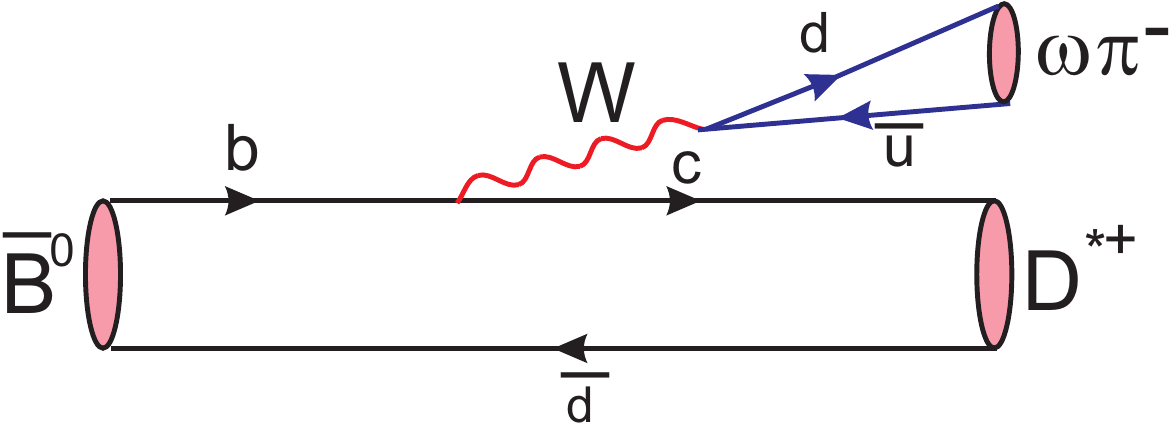}  \\
{\bf (a)} & {\bf (b)} \\
\end{tabular}
\caption{(color online). (a) Color-suppressed and (b) color-favored quark
diagrams taken from ref.~\cite{belle} and showing $D^{**}$ and $\omega\pi$ 
production in 
$\bar{B}^0 \to D^{*+} \omega \pi^-$ decays.}
\label{fig:decay}
\end{center}
\end{figure*}

In 2015 the Belle collaboration reported a detailed amplitude analysis of the 
$\bar{B}^0 \to D^{*+} \omega \pi^-$ decays based on the full $B\bar{B}$ data 
sample at the $\Upsilon(4S)$ resonance~\cite{belle}.  
This is a clean enough system to perform a factorization test because background 
associated with the $\omega$ meson emitted from the $(\bar{c}b)$ current is 
suppressed.
The decay under consideration arises predominantly due 
to the weak interaction with color-favored and color-suppressed contributions 
shown in figure~\ref{fig:decay} taken from ref.~\cite{belle}.

The $D^{**}$ and $\rho$-resonant contributions are fully extracted from 
amplitude analysis in ref.~\cite{belle}. It allows us to test factorization 
separately in their production regions.

The $D^{**}$ rate is found to be of the order of $15\%$~\cite{belle}. A 
test of factorization in the $D^{**}$ region is based on the polarization 
measurements of the $D^{**}$ resonances.
Observable significant transverse polarizations together with heavy quark 
symmetry can imply nonfactorizable effects in this channel. The factorization 
and structure of the $(V-A)$ weak current forbids production of resonances 
with spin $>$ 1. Observation of the  $D^*_2(2460)$ resonance with spin of 2
reported in ref.~\cite{belle} therefore directly violates factorization in this region. 

Our goal is to test factorization in a color-favored channel with production 
of two vector resonances, off-shell $\rho(770)$ and $\rho(1450)$, which 
dominate in the total branching fraction.
The influence of the model parameters used to describe the $D^{**}$ states on the accuracy of the
factorization test in the $\omega\pi$ resonant region is not significant. 
This is demonstrated by table III in ref.~\cite{belle} where the dominant model 
uncertainties are shown. The last column in this table shows the uncertainty 
from the mixing of the $D^{**}$ states. Variations of parameters of the 
$\rho$-like states, associated with this uncertainty, are small compared 
to other uncertainties. 

A test of factorization can be performed in two ways. 
The first utilizes the fraction of the longitudinal polarization 
$\mathcal{P}_{D^*}$ of the $D^*$ which should be the same as in the related 
semileptonic decay $\bar{B}^0 \to D^{*+} l^- \nu_l$ at squared four-momentum 
transfer equal to the mass squared of the intermediate $\rho$-like 
resonance~\cite{goldstein}. Such a test has been performed in the 
CLEO~\cite{cleo} and BaBar~\cite{babar} analyses and confirmed the 
factorization validity within their experimental accuracy. In the Belle 
amplitude analysis~\cite{belle} the longitudinal polarization 
$\mathcal{P}_{D^*}$ is fixed in part from the factorization prediction. The 
relative normalizations of the helicity amplitudes are fixed at values measured
in $\bar{B}^0 \to D^{*+} l^- \nu_l$~\cite{babar2}. Free mass and width of the 
$\rho(1450)$ can slightly affect the $\mathcal{P}_{D^*}$ value but it agrees 
well with the factorization prediction. Such a test is sensitive only to the 
corrections affecting different partial waves. It is also a local test at a 
given resonance mass point not considering the dynamic behaviour of 
corrections to factorization. 

In the other test a form factor of $\omega\pi$ production in the 
hadronic $\bar{B}^0 \to D^{*+} \omega \pi^-$ decay is studied. In the 
factorization approximation this form factor should be the same as in 
$\tau \to \omega \pi \nu_{\tau}$ decays. Assuming the vector current to be the 
same in electromagnetic and weak decays (conservation of vector current 
or CVC), 
similar correspondence should also exist for the $e^+e^- \to \omega \pi^0$ 
process. In such a case, the isovector part of the electromagnetic current 
$J^{\rm el}_{\mu}$ matches the weak charged current:
\begin{equation} 
<\rho^0|J^{\rm el}_{\mu}|0>\,=\,\frac{1}{\sqrt{2}}<\rho^-|\bar{u}\gamma_{\mu}d|0>{.} 
\label{eq:cvc}
\end{equation} 
Such a test uses the distribution of the invariant mass squared, 
$M^2(\omega\pi)$, and allows us to test factorization over the whole accessible 
kinematic range. Transition form factors $F^{\tau}_{\omega\pi}(q^2)$ in 
$\tau \to \omega \pi \nu_{\tau}$ decays and $F^{e^+e^-}_{\omega\pi}(q^2)$ in $e^+e^- \to \omega \pi^0$ processes can be measured directly from the differential width and the Born cross section. The transition form factor $F^B_{\omega\pi}(q^2)$ 
can be evaluated from the amplitude analysis of the 
$\bar{B}^0 \to D^{*+} \omega \pi^-$ decays.

The signal matrix element $M_{\rm sig}$ 
for a color-favored channel determines the $\omega\pi$ transition form factor 
$F^B_{\omega\pi}(q^2)$ as a function of $q^2 = M^2(\omega\pi)$:
\begin{equation}
M_{\rm sig}\,=\,\frac{G_F}{\sqrt{2}}V^{}_{\rm cb}V^{*}_{\rm ud}a_1F^B_{\omega\pi}(q^2)\epsilon^{\mu\nu\alpha\beta}J^{(B\to D^{*})}_{\mu}v^{*}_{\nu}q_{\alpha}p_{\beta}{.}
\label{eq:mrho}
\end{equation}  
Here, $a_1$ is the relevant QCD coefficient, $J^{(B\to D^*)}_{\mu}$ describes a 
transition current of $B \to D^*$, $v_{\mu}$ is a four-vector of the $\omega$ 
meson polarization, $q_{\mu}$ is a four-momentum of the $\omega\pi$ pair 
and $p_{\mu}$ is a four-momentum of the $\omega$. 
In eq.~\eqref{eq:mrho}, the nonfactorizable corrections are encoded into the 
$a_1F^B_{\omega\pi}(q^2)$ product as well as in the $J^{(B\to D^*)}_{\mu}$ current. 
If factorization is exact, we have for all available values of $q^2$:
\begin{equation}
    a_1 F^B_{\omega\pi}(q^2)\,=\,\left(c_1(\mu)+\frac{c_2(\mu)}{3}\right)F^{\tau (e^+e^-)}_{\omega\pi}(q^2){,}
    \label{eq:fact}
\end{equation}
 where the Wilson coefficients $c_1(\mu)$ and $c_2(\mu)$ are renormalized 
at the scale of $\mu$, and the current $J^{(B\to D^*)}_{\mu}$ is extracted from 
$B \to D^* l \nu_l$ data. Results of the amplitude analysis in ref.~\cite{belle} 
are obtained under the assumption that corrections affecting the polarization 
of the $D^*$ are absent. In such a case, the current $J^{(B\to D^*)}_{\mu}$ takes 
into account only nontrivial final-state interaction phases in helicity 
amplitudes (see Appendix C in ref.~\cite{belle}) which cancel after 
integration over angular variables. These phases have been measured in 
ref.~\cite{belle}, although uncertainties are large.
 
As a consequence, such a test is not exhaustive and should be considered as a 
complementary one to the first test discussed above. 
The corrections to factorization $\delta_{\rm NF}$ in frame of the discussed 
test are determined as
\begin{equation}
    1+\delta_{\rm NF}\,=\,\frac{|a_1 F^B_{\omega\pi}(q^2)|} {\left(c_1(\mu)+\frac{c_2(\mu)}{3}\right)|F^{\tau (e^+e^-)}_{\omega\pi}(q^2)|}{.}
\label{eq:factcor}
\end{equation}

The parameter $\delta_{\rm NF}$ in eq.~\eqref{eq:factcor} describes 
nonfactorizable contributions to $B \to D^* \omega \pi$ decay appearing in 
different ways. 
The first way is related to the $1/N_c$ expansion in QCD where factorization 
does not depend on the mass of the produced $\omega\pi$ 
system~\cite{NCexpansion}. As factorization ignores the color of the quarks 
produced by the virtual W, it is instructive to rewrite the $B$ decay 
amplitude in a way that restores this dependence. In this case the effective 
coefficient $a_1$ in eq.~(1.2) is written as:
\begin{equation}
a_1\,=\,\left(c_1(\mu)+\frac{c_2(\mu)}{3}\right)[1+\epsilon_1(\mu)]+c_2(\mu)\epsilon_8(\mu){,}
\end{equation}
where hadronic parameters $\epsilon_1(\mu)$ and $\epsilon_8(\mu)$ determine 
the nonfactorizable contributions appearing from the color-singlet and 
color-octet current operators. In the large $N_c$ limit, $|\epsilon_1|\ll 1$, 
whereas contributions from $\epsilon_8$ can be more sizeable and 
$\epsilon_8>0$~\cite{NCexpansion}. In such a case,
\begin{equation}
a_1\,=\,c_1(\mu)+\zeta c_2(\mu){,}
\end{equation}
where $\zeta=1/3+\epsilon_8(\mu)$. Since $c_2(\mu)<0$, the effective constant 
$a_1$ is expected to be slightly less than the naive factorization prediction.
    
The second way corresponds to the perturbative QCD~\cite{pqcd}. This picture 
of factorization expands the $B$ decay amplitude in powers of 
$M(\omega\pi)/m_b$. As the b-quark is heavy, the corrections to factorization 
are suppressed for a light $\omega\pi$ system. As the mass of the 
$\omega\pi$ system increases, the corrections become significant. It should 
be seen as a difference between form factor shapes extracted from $B$ data 
and $\tau$ ($e^+e^-$) data when $M(\omega\pi)$ increases.  

The factorization test could be applied above the $\omega \pi$ threshold. The 
form factor shape in the low-energy region is evaluated by extrapolation from 
the $\omega\pi$ production region and direct measurements in the $\omega$ 
conversion decay region. More precise data in the conversion region were 
obtained by the NA60 collaboration~\cite{na60} from a study of the 
$\omega \to \mu^+ \mu^- \pi^0$ decay. The NA60 data lie strongly above the 
prediction of the vector meson dominance (VMD) model which quite well describes 
the form factor in $e^+e^-$ annihilation.

\section{Analysis of the $F_{\omega\pi}(q^2)$ form factor}

The form factor $F^{B}_{\omega\pi}(q^2)$ of $\omega \pi$ production in 
$B \to D^* \omega \pi$ decays can be defined as
\begin{equation}
F^{B}_{\omega\pi}(q^2)\,=\,\tilde{g}f_{\omega\pi}(q^2){,}
\label{eq:ffbelle0}
\end{equation}
where $\tilde{g}$ is a coupling constant calculated from the combined fraction 
$f_{\rho+\rho'}$ of the $\rho$ and $\rho'$ in the total branching fraction 
$B \to D^* \omega \pi$ and
\begin{equation}
f_{\omega\pi}(q^2)\,=\,\sqrt{q^2}\left(\frac{F_{\rho}(q^2)}{D_{\rho}(q^2)}+A_{\rho'} e^{i \phi_{\rho'}} \frac{F_{\rho'}(q^2)}{D_{\rho'}(q^2)}\right){.}
\label{eq:ffbelle}
\end{equation}  
In eq.~\eqref{eq:ffbelle}, $F_{\rho}(q^2)$ ($F_{\rho'}(q^2)$) is the 
$\rho(770)$ ($\rho(1450)$) form factor in the decay to the $\omega\pi$ final 
state and $D_{\rho}(q^2)$ ($D_{\rho'}(q^2)$) is the Breit-Wigner denominator 
describing the $\rho(770)$ ($\rho(1450)$) shape. 
The form factors $F_{\rho}(q^2)$ and $F_{\rho'}(q^2)$ restrict a rapid growth 
of the $B$ decay amplitude with $p_{\omega}$ which is the magnitude of the 
$\omega$ three-momentum in the $\omega\pi$ rest frame. The simple 
Blatt-Weisskopf parameterization is used for them:
\begin{equation}
F_{\rho}(q^2)\,=\,\frac{1}{1+(rp_{\omega})^2}{,} \,\,\,
F_{\rho'}(q^2)\,=\,\sqrt{\frac{1+(rp_{0,\omega})^2}{1+(rp_{\omega})^2}} {,}
\label{eq:ffrho}
\end{equation}
where $r=1.6$ ${\rm GeV}^{-1}$ is a typical hadronic scale and
$p_{0,\omega}$ is the $\omega$ three-momentum $p_{\omega}$, when $q^2=m^2_{\rho'}$.
The functions $D_{\rho(\rho')}(q^2)$ are given by 
\begin{equation}
D_{\rho(\rho')}(q^2)\,=\,q^2-m^2_{\rho(\rho')} + i \sqrt{q^2}\Gamma_{\rho(\rho')}(q^2){,}
\label{eq:bw}
\end{equation}
where $\Gamma_{\rho}(q^2)$ and $\Gamma_{\rho'}(q^2)$ are the $q^2$-dependent 
widths of the $\rho(770)$ and $\rho(1450)$ resonances. 
The width $\Gamma_{\rho}(q^2)$ ($\Gamma_{\rho'}(q^2)$) is defined in 
ref.~\cite{belle} (eqs.~(B5) and~(B6)) with the additional factor of 
$m_{\rho}/\sqrt{q^2}$ ($m_{\rho'}/\sqrt{q^2}$) arising from the different 
definitions of $D_{\rho(\rho')}(q^2)$ in eq.~\eqref{eq:bw} and 
in ref.~\cite{belle} (eq.~(B3)).
In eq.~\eqref{eq:ffbelle}, parameterization of the $f_{\omega\pi}(q^2)$ form 
factor is different from the VMD model. The difference is that the resonance 
masses $m_{\rho}$ and $m_{\rho'}$ are replaced with the invariant mass 
$M(\omega\pi)$. It was found that a fit to the $B$ data with the VMD function 
leads to the worse data description. It corresponds to the negative 
log-likelihood value lying about $3\sigma$ away from the global minimum 
obtained with the model in eq.~\eqref{eq:ffbelle}. Therefore the model in 
eq.~\eqref{eq:ffbelle} is chosen as acceptable for the current $B$ data 
description.

To obtain the form factor defined in eq.~\eqref{eq:ffbelle0}, a coupling 
constant $\tilde{g}$ should be determined but it could be extracted 
only combined with the coefficient $a_1$. 
A product of $a_1\tilde{g}$ is determined by the color-favored branching 
fraction:
\begin{equation}
a_1\tilde{g}\,=\,\frac{8\pi\sqrt{3\pi}m_B}{G_F\mathcal{F}(1)|V_{\rm cb}||V_{\rm ud}|}\sqrt{\frac{f_{\rho+\rho'}\Gamma(B \to D^* \omega \pi)}{J}}{,} 
\label{eq:g}
\end{equation}
where 
\begin{equation}
J\,=\,\int p^3_{\omega}p_{D^*}\sqrt{q^2}|f_{\omega\pi }(q^2)|^2(|f_S(q^2)|^2+|f_P(q^2)|^2+|f_D(q^2)|^2)dq^2{.}
\label{eq:J}
\end{equation}
In eq.~\eqref{eq:J}, $f_S(q^2)$, $f_P(q^2)$ and $f_D(q^2)$ are the partial wave 
form factors describing a transition $B \to D^*$. In the frame of heavy quark 
effective theory they can be related to the Isgur-Wise function with a 
parameter $\rho^2$, relative factors $R_1$ and $R_2$ and normalization factor 
$\mathcal{F}(1)$. The linear approximation of the Isgur-Wise function is used 
in ref.~\cite{belle} with parameters $\rho^2$, $R_1$ and $R_2$ measured in 
ref.~\cite{babar2}. This very simple parameterization was sufficient for 
amplitude analysis in ref.~\cite{belle}. However, at the moment we have to 
know the product $\mathcal{F}(1)\times |V_{\rm cb}|$ which has been measured 
with the best accuracy by the Belle collaboration~\cite{waheed} in the 
Caprini-Lellouch-Neubert (CLN) parameterization~\cite{cln}. For consistency, 
we refit the $\bar{B}^0 \to D^{*+} \omega \pi^-$ data~\cite{belle} with the CLN 
function where parameters $\rho^2$, $R_1$ and $R_2$ are fixed at their values 
from ref.~\cite{waheed}. This fit results in a relative $\rho'$-strength 
$A_{\rho'}=0.19 \pm 0.05$, relative $\rho'$-phase $\phi_{\rho'}=2.52 \pm 0.11$ 
rad, $\rho'$-mass $m_{\rho'}=1540 \pm 22$ MeV and $\rho'$-width 
$\Gamma_{\rho'}=304\pm 49$ MeV, where a statistical error only is shown.
Finally, using the product $\mathcal{F}(1)\times |V_{\rm cb}|$ fixed at the 
value $(35.06 \pm 0.58)\times 10^{-3}$~\cite{waheed}, $|V_{\rm ud}|=0.9742 \pm 0.0002$~\cite{pdg} as well as the product $f_{\rho+\rho'}\times\mathcal{B}(B\to D^* \omega\pi)=(1.90 \pm 0.17)\times 10^{-3}$ and $\Gamma_B=4.326\times 10^{-13}$ GeV~\cite{pdg}, we obtain $a_1 \tilde{g}=2.66 \pm 0.28$. The error of 
$a_1\tilde{g}$ is calculated from the error propagation formula. The main 
source of the error is related to the integral calculation $J$, where the 
covariance matrix of model parameters is applied. 

In the factorization approximation, the effective coefficient $a_1$ at next-to-leading 
order was obtained in ref.~\cite{beneke}. It is renormalized at the scale of 
$b$-quark mass $\mu=m_b$ and leads to the value of $a_1(m_b)=1.02$. The uncertainty of 
this value is related to the scale where Wilson coefficients are evaluated. 
Taking into account the values of $a_1$ calculated at the scales $\mu=2m_b$ 
and $\mu=m_b/2$ (see~\cite{beneke}), we obtain $a_1=1.02 \pm 0.02$.
In other words, $\tilde{g}=2.61 \pm 0.28$. The coupling $\tilde{g}$ is the 
product of the $\rho$-meson weak decay constant $f_{\rho}$ and 
$g_{\rho\omega\pi}$ coupling for the $\rho \to \omega \pi$ transition.
The value for $\tilde{g}$ can be compared with that measured from the 
$e^+e^-$ data.
The combined SND2000~\cite{snd2000} and SND2016~\cite{snd2016} data are fit 
by the model in eq.~\eqref{eq:ffbelle0}, which is used for the $B$ data. The 
mass and width of the $\rho'$ resonance as well as the $\tilde{g}$ constant 
are free parameters in the fit which gives $\tilde{g}=3.05 \pm 0.02$. Taking 
into account that $a_1\tilde{g}=2.66 \pm 0.28$, we obtain $a_1=0.87 \pm 0.09$. 
This value is less than $a_1(m_b)=1.02 \pm 0.02$ obtained when the 
nonfactorizable corrections are neglected but they are consistent with each 
other within the statistical accuracy.

The $g_{\rho\omega\pi}$ value depends on parameterization of the $F_{\rho}(q^2)$ 
form factor in eq.~\eqref{eq:ffrho} and should be determined from 
$\tilde{g}$ and $f_{\rho}$.
We can estimate $f_{\rho}$ experimentally from the decay width of 
$\rho \to e^+ e^-$ using the CVC relation in eq.~\eqref{eq:cvc}
\begin{equation}
f_{\rho}\,=\,\sqrt{\frac{3m_{\rho}\Gamma(\rho \to e^+ e^-)}{2\pi \alpha^2}}{.}
\end{equation}
It gives $f_{\rho}=(0.220 \pm 0.001)$ GeV. Finally, we obtain 
$g_{\rho\omega\pi}=(11.9 \pm 1.3)$ GeV${}^{-1}$.
This value can be compared with the SND value of $g_{\rho\omega\pi}=(13.9 \pm 0.1)$ GeV${}^{-1}$ measured with the same $F_{\rho}(q^2)$ form factor. The difference 
between these values is not statistically significant. If the hadronic parameter $r \to 0$ in eq.~\eqref{eq:ffrho}, the SND measurement gives $g_{\rho\omega\pi}=(15.9 \pm 0.4)$ GeV${}^{-1}$. This value is in good agreement with the prediction of
QCD sum rules of $16$ GeV${}^{-1}$~\cite{lublinsky}.

Now, when the form factor $F^B_{\omega\pi}(q^2)$ is fully extracted from the $B$ 
data, it can be compared to $\tau$ and $e^+e^-$ data in the frame of 
factorization. An additional factor $\sqrt{2}$ must be used to convert from 
the electromagnetic form factor to the weak one using the CVC relation 
in eq.~\eqref{eq:cvc}. 
 
\begin{figure*}[htb!]
\begin{center}
\includegraphics[width=\textwidth]{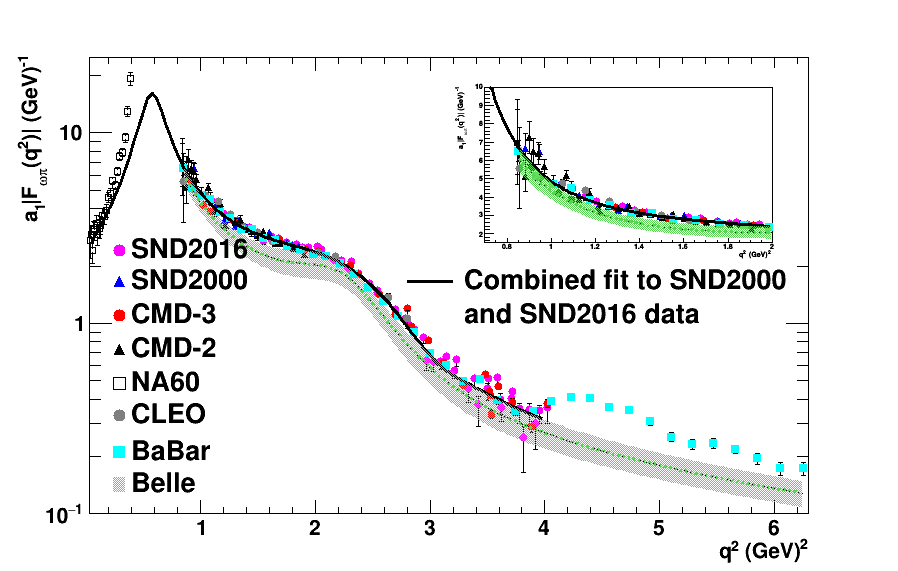} 
\caption{(color online). The $\omega \pi$ production form factor 
weighted by the value of $a_1$ with $a_1=1.02 \pm 0.02$~\cite{beneke} and 
shown in a log scale. The points with error bars show $e^+e^-$ data by the 
SND in ref.~\cite{snd2000} (blue triangles, SND2000) and
ref.~\cite{snd2016} (magenta circles, SND2016), 
CMD-3 in ref.~\cite{cmd3} (red circles), CMD-2 in ref.~\cite{cmd2} (black triangles) 
and BaBar collaborations in ref.~\cite{babar2017} (cyan squares) as well 
as $\omega$ conversion data by the NA60 in ref.~\cite{na60} (open squares) and 
$\tau$ lepton data by the CLEO collaborations in ref.~\cite{cleo_tau} 
(grey circles). The 
black solid line is the combined fit result to the SND2000 and SND2016 data 
taken from ref.~\cite{snd2016}. 
The green dotted line corresponds to the function 
$a_1/\sqrt{2}|F^{B}_{\omega\pi}(q^2)|$. The green dashed area represents the 
$68\%$ confidence level contour of statistical uncertainties of the model 
parameters in eq.~\eqref{eq:ffbelle0}. The zoomed-in plot in the low $q^2$ 
region is also shown.}
\label{fig:result}
\end{center}
\end{figure*}

Figure~\ref{fig:result} demonstrates experimental data of $a_1|F_{\omega\pi}(q^2)|$ as a function of $q^2$ measured in $\tau$ lepton decays to $\omega \pi \nu_{\tau}$ by the CLEO collaboration (grey circles)~\cite{cleo_tau} and 
$e^+e^- \to \omega \pi^0$ processes with the subsequent decay of the $\omega$ 
to $\pi^+\pi^-\pi^0$ or $\pi^0 \gamma$.

The data from $e^+e^-$ collisions were obtained either 
in direct $e^+e^-$ annihilation by the CMD-2 (black triangles)~\cite{cmd2}, 
CMD-3 (red circles)~\cite{cmd3} as well as SND 
(blue triangles~\cite{snd2000} and magenta circles~\cite{snd2016}) 
collaborations or using initial-state radiation (ISR) by the BaBar 
collaboration (cyan squares)~\cite{babar2017}. The NA60 data 
(open squares)~\cite{na60} in the conversion $\omega \to \pi^0 \mu^+ \mu^-$ 
decays are also shown at low $q^2$ values.
In figure~\ref{fig:result}, the green dotted line shows the product 
$a_1|F^B_{\omega\pi}(q^2)|/\sqrt{2}$ obtained above from the $B$ decay data. The dashed 
area corresponds to $\pm 1\sigma$ deviation from the line taking into account 
the statistical covariance matrix of parameters. The solid black line shows 
the fit result to the SND data (\cite{snd2000},\cite{snd2016}) by the VMD model. 
The VMD model cannot simultaneously describe the $e^+e^-$ and $\omega$ 
conversion data. 

The uncertainties in the parameterization of the form factor in 
eq.~\eqref{eq:ffbelle0} give the model error which is not shown in figure~\ref{fig:result}. The reason is that the statistical uncertainties are still large. 
We believe that the model uncertainty distorting the form factor shape will 
decrease with a statistical error.

The $\omega\pi$ form factor extracted from the Belle data set in 
ref.~\cite{belle} agrees well in shape with predictions obtained from 
$\tau$ and $e^+e^-$ data in the region $q^2 < 4$ GeV${}^2$ but has a lower 
normalization. However, the difference between normalizations is not 
significant and both values are consistent with each other within the Belle 
form factor accuracy.
More $B$ data are needed to make a detailed comparison.

\begin{figure*}[htb!]
\begin{center}
\includegraphics[width=\textwidth]{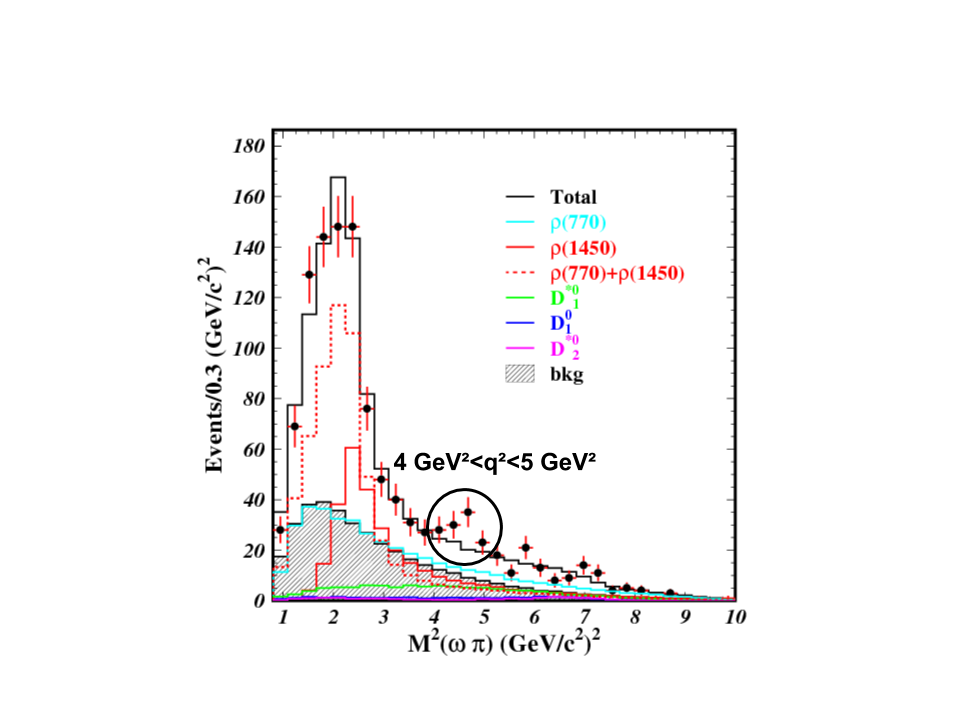} 
\caption{(color online). The distribution of $q^2$ for $B\to D^* \omega \pi$ events 
measured by the Belle collaboration and taken from ref.~\cite{belle}. 
Points with error 
bars show experimental data, histograms in color represent several resonant 
contributions and black histogram describes the signal fit. The region 
$4$ GeV${}^2$ $<$ $q^2$ $<$ $5$ GeV${}^2$ is specially contoured.}
\label{fig:belle}
\end{center}
\end{figure*}

In the region above $q^2 = 4$ GeV${}^2$ the $e^+e^-$ ISR data collected 
by the BaBar collaboration~\cite{babar2017} are available in the $e^+e^-$
sector. The clear bump is 
seen in the region between $q^2=4$ GeV${}^2$ and $q^2=5$ GeV${}^2$. 
This bump can be also seen in the 
Belle data. Figure~\ref{fig:belle} shows the overall $q^2=M^2(\omega\pi)$ 
distribution for the $B \to D^* \omega \pi$ decays. The region of interest 
$4$ GeV${}^2$ $<$ $q^2$ $<$ $5$ GeV${}^2$ is specially contoured. 
A similar structure is seen here but it is not statistically significant. 
The black histogram describes the data using the signal model in 
ref.~\cite{belle}. 
The signal model is not sensitive to the bump and has a smooth shape.
Note that the bump between $q^2=4$ GeV$^2$ and 5 GeV$^2$ is most probably
due to the $\rho(2150)$, a broad structure around $2.15-2.2$ GeV observed
in different final states~\cite{pdg}. 
It could be a reason of different shapes of the $\omega \pi$ form factor 
extracted in $e^+e^-$ and $B$ decay data shown in figure~\ref{fig:result}. 
The Belle sensitivity is not sufficient to identify the bump in 
figure~\ref{fig:belle}.  
More $B$ data are needed to make a definite conclusion about its origin. 

\section{Conclusion}
Corrections to factorization are studied in $B\to D^* \omega \pi$ decays 
through 
the $\omega\pi$ production form factor extracted from the amplitude analysis 
of the Belle data~\cite{belle}. The form factor is compared to available 
measurements of $\tau$-lepton decays and $e^+e^-$ annihilation. 

The advantage of this approach is that the contributions of $D^{**}$ states 
and $\rho$-like resonances can be exactly separated from each other. It is 
crucial to perform the factorization test in the $\rho$-like production region 
because the integrated contribution from the $D^{**}$ states was found to be 
significant (about $15\%$). The $q^2=M^2(\omega\pi)$ differential mass 
distribution in the $\rho$-like enriched region deteriorates by the 
$D^{**}$ contribution and the factorization interpretation for the decay 
becomes less clear. It prevents from using the experimental $q^2$ distribution
normalized to the semileptonic rate. 
Our approach is model-dependent but the model uncertainty will be controlled in 
a more rigorous way when the statistical accuracy is improved. 

Figure~\ref{fig:result} shows a product $a_1|F_{\omega\pi}(q^2)|$ of the 
$\omega \pi$ form factor and efficient constant $a_1$ obtained from 
the $B$ data and predicted from factorization using the $\tau$ and $e^+e^-$ 
data. 
The large uncertainties appearing when the $\omega\pi$ form factor is 
extracted from the $B$ data do not allow us to observe corrections to 
factorization in a statistically significant manner. 
Experimental analysis of available data demonstrates 
that $\omega \pi$ production is similar in $B$ decays and 
$e^+e^-$/$\tau$-lepton processes.
The uncertainty in the range below $q^2=1.2$ GeV (around 10\%) is mainly due 
to the knowledge of the $a_1\tilde{g}$ product whereas parameters of the 
$\rho'$ resonance contribute to the uncertainty at higher $q^2$ values 
(around 15\%).

For a precision study of corrections to factorization in the 
$\bar{B}^0 \to D^{*+} \omega \pi^-$ decay it is necessary to perform a 
high-statistics analysis with a data set available at LHCb and, in future, 
Belle II detectors.
 
The authors thank A.~Khodjamirian and A.~Kuzmin for useful discussions. 
SE is grateful
to Munich Institute for Astroparticle and Particle Physics, where part of
this work has been done.
The work is supported by the Grant of the Russian Federation Government, 
Agreement No.~14.W03.31.0026.

\end{document}